# Multifactorial cancer treatment outcome prediction through multifaceted radiomics

Zhiguo Zhou[1], David Sher[1], Qiongwen Zhang[2], Pingkun Yan[3], Jennifer Shah[1], Nhat-Long Pham[1], Michael Folkert[1], Steve Jiang[1], and Jing Wang[1, *]

[1]Medical Artificial Intelligence and Automation Laboratory (MAIA Lab), Department of Radiation Oncology, UT Southwestern Medical Center, Dallas, TX

[2]State Key Lab of Biotherapy and Cancer Center, West China Hospital, Sichuan University, Chengdu, China

[3]Department of Biomedical Engineering, Rensselaer Polytechnic Institute, Troy, NY

*Abstract--*Accurately predicting the treatment outcome plays a greatly important role in tailoring and adapting a treatment planning in cancer therapy. Although the development of different modalities and personalized medicine can greatly improve the accuracy of outcome prediction, they also bring the three mainly simultaneous challenges including multi-modality, multi-classifier and multi-criteria, which are summarized as multifactorial outcome prediction (MFOP) in this paper. Compared with traditional outcome prediction, MFOP is a more generalized problem. To handle this novel problem, based on the recent proposed radiomics, we propose a new unified framework termed as multifaceted radiomics (M-radiomics). M-radiomics trains multiple modality-specific classifiers first and then optimally combines the output from the outputs of different classifiers which are trained according to multiple different criteria such as sensitivity and specificity. It considers multi-modality, multi-classifier and multi-criteria into a unified framework, which makes the prediction more accurate. Furthermore, to obtain the more reliable predictive performance which is to maximize the similarity between predicted output and labelled vector, a new validation set based reliable fusion (VRF) strategy and reliable optimization models as well as a new recursive two stage hybrid optimization algorithm (RTSH) were also developed. Two clinical problems for predicting distant metastasis and locoregional recurrence in head & neck cancer were investigated to validate the performance and reliability of the proposed M-radiomics. By using the proposed RF strategy and RTSH optimization algorithm, the experimental results demonstrated that M-radiomics performed better than current radiomic models that rely on a single objective, modality or classifier.

## I. Introduction

Outcome prediction in cancer treatment refers to staging, tumor response to therapy, rates of local recurrence, evolution to metastatic disease, development of toxicity during follow up, or a combination of these endpoints [1]. Accurately predicting outcomes prior to or even during therapy is of great value, as it facilitates more effective treatment planning for individual patients [2]. For example, for patients in early stage non-small cell lung cancer, accurately predicting distant failure after stereotactic body radiation therapy (SBRT) may adopt additional systemic therapy so as to improve the overall survival [3].

Similarly, when receiving external beam radiation therapy with concurrent chemotherapy and intracavitary brachytherapy, at least 20% patients with locally advanced cervical cancer still develop distant failure [4]. As such, it is essential to predict distant failure for patients with high risk so that better treatment outcomes with intensified treatment modalities can be achieved. In head & neck cancer, since one third of the body's lymph nodes are located in the head and neck, lymph node metastases has become one of the most important prognostic factor and early prediction is necessary for treatment optimization [5]. As such, building an accurate and reliable treatment outcome prediction model is of great importance in cancer care.

With the development of modern diagnostic as well as treatment modalities, and a great progress in personalized medicine has, however, created new three important challenges simultaneously for predicting treatment outcome. First of all, different source of information are always available for building predictive model, e.g., different imaging modalities such as PET, CT and MRI. However, the challenge is how to optimally integrate available and diverse multimodal information in a quantitative way so that the reliable and accurate outcomes can be provided. For example, FDG-PET scanning measures glucose metabolism, while CT scanning provides attenuation coefficient information for x-rays. So a simple combination of the features extracted from these different modalities may not yield an optimal prediction. Secondly, the progress of artificial intelligence and machine learning has provided multiple choices for model construction. However, different techniques (classifier) are always associated with distinct inherent limitations and the choice of modelling technique has been shown to affect the prediction performance. Therefore, the challenge is how to choose an "optimal" or a "preferred" classifier for a particular application [6]. Thirdly, it is always difficult to make balance between sensitivity and specificity especially when positive and negative cases are imbalanced in training dataset [3]. For example, when the number of patients with distant failure in lung SBRT is greatly lower than the patients without distant failure, the sensitivity will be low even though the accuracy may be high as it is used as the objective function. So the challenge is how to balance the two objects (sensitivity and specificity).

The above three challenges can summarized as multi-modality, multi-classifier and multi-criteria. Due to the importance of the three challenges and to distinguish from the traditional outcome prediction problem, **a new problem named as multifactorial outcome prediction (MFOP) is introduced for describing the outcome prediction with three challenges for the first time.** The aim of MFOP is to get more accurate as well as reliable and interpretable predictive results. Compared with the traditional outcome prediction, MFOP is a more generalized problem framework.

With the recent advances in medical imaging technology, radiomics, referring to extracting and analyzing a huge number of quantitative image features [7-10], provides a unprecedented opportunity to improve personalized treatment assessment [11]. In a recent study, radiomic features were extracted from CT images in [12] to predict survival time in NSCLC with an accuracy of 77.5%. When CT based radiomics features were combined with the clinical model to predict distant metastasis in lung adenocarcinoma, the performance was significantly improved [13]. By selecting radiomic features from FDG-PET images in [2], the accuracy for predicting lung tumor recurrence can achieve 0.94. Although it is demonstrated that radiomics has achieved great success for handling traditional outcome prediction problem, only a few studies investigated the three challenges in MFOP. The authors proposed the multi-objective radiomics model which considers both sensitivity and specificity as the objective functions simultaneously [3], and the predictive results greatly improved compared with single objective based radiomics model. Then multi-modality/classifier radiomics model were also investigated, and the case studies showed that combining multiple modalities or different classifiers in a reasonable way can improve accuracy and reliability of the outcome [6]. However, there has been no a unified model which can deal with the three challenges simultaneously in MFOP so far.

In this work, **a unified framework termed as multifaceted radiomics (M-radiomics) is proposed for handling MFOP for the first time**. M-radiomics consists of three steps: (1) Multimodal image segmentation; (2) Multi-criterion feature extraction and selection; (3) M3 (multi-modality, multi-classifier and multi-objective) predictive model construction. Since the tumor locations are same in different modalities, it is possible that the tumors are segmented in a collaborative way at first step. In the second stage, as the optimal feature subset are not only determined by relevance and redundancy, but also based on the predictive performance including sensitivity and specificity, they should be chosen in multi-criterion way. To overcome the limitation of the conventional single-objective model, M-radiomics considers both sensitivity and specificity as objective functions during model training. Instead of choosing a preferred classifier or blindly combining features extracted from different modalities, M-radiomics trains modality-specific classifiers first and then optimally combines the output from the output of multiple modality-specific classifiers that is trained according to multiple different criteria such as sensitivity and specificity. As such, the three challenges associated with current radiomics are solved through multi-modality, multi-classifier and multi-criteria in the proposed M-radiomics.

Furthermore, the information extracted from different sources (e.g., modality and classifier) needs to be combined to yield a final prediction result in M3 predictive model. Originally designed to combine information from different classifiers, the classifier fusion strategy offers an effective solution for both multi-modality and multi-classifier models [14-16]. To obtain more reliable predictive results, we propose a validating set based reliable fusion (VRF) strategy that not only considers the relative importance between different classifiers, but also considers the reliability of the classifier itself. In VRF, reliability of the individual classifier output is first determined by considering the output probabilities from the validation set. If the output probability of one classifier is similar to most output probabilities in validation set, the reliability of this classifier should be high. Both reliability and weight are combined with output probabilities of individual classifiers to generate a final output score by an analytical evidential reasoning rule [6, 17]. On the other hand, when training the relative weight and the parameters in individual classifiers, not only sensitivity and specificity are considered as the objective function, but also two new objectives which are similarity based sensitivity and specificity are introduced so as to achieve more reliable results. Hence, a new reliable optimization model is proposed. Correspondingly, based on our previous work [3], a new recursive two stage hybrid optimization algorithm (RTSH) is proposed in this work. In the first stage, iterative multi-objective immune algorithm [3] is performed for maximizing sensitivity and specificity, and similarity based multi-objective immune algorithm is designed for maximizing similarity based sensitivity and specificity in the second stage. These two stages are performed recursively until the algorithm achieves the maximal generation.

Two clinically significant problems in head & neck (H&N) cancer are selected to validate the performance and reliability of the proposed M-radiomics model. The first problem is predicting distant metastasis. Although the locoregional control of H&N cancer after radiotherapy in early stage has been improved, the developments of distant metastasis is the leading causes of treatment failure and death [18] [19], ranging from 6.1% to 16.3% [20]. For patients at high-risk of distant metastasis after definitive treatment, intensification with immediate systemic therapy may reduce the risk of distant relapse and improve overall survival. Therefore, accurately predicting of patients with high-risk distant metastasis in early stage is central for to improve treatment outcome of H&N cancer patients. Standard of care medical images such as 18F-fluorodeoxyglucose positron emission tomography (FDG-PET) and X-ray computed tomography (CT) can carry the immense source of potential data for decoding the tumor phenotype [7]. When jointly using FDG-PET and CT, it has the unique capability to image metabolically active lesions and provide more anatomical details than only PET [21]. The second problem taken up here is to predict locoregional recurrence after radiation therapy. Although a great progress has been made in treating H&N cancer, a substantial number of patients occur locoregional recurrence within the first three years [22]. Early prediction of locoregional recurrence is crucial to improve treatment outcome of these high-risk patients by potentially early salvage treatment. Similar with the first problem, FDG-PET and CT also have been considered as a source of prognostic biomarkers for locoregional recurrence prediction. In this paper, we exploit the improved and reliable predictive results for the two typical MFOP problems based on FDG-PET and CT images through M-radiomics.

The reminder of the paper is organized as follows. The problem description for multifactorial outcome prediction is described in Section II. Section III describes the proposed M-radiomics model, which contains overview, validation set based reliable fusion strategy, reliable optimization model as well as recursive two stage hybrid optimization algorithm, training and testing procedures based on M-radiomics. Section IV presents the experimental results and analysis for the two prediction problems. The discussion is presented in section V and the conclusions are given in section VI.

## II. Problem description

Since multifactorial outcome prediction (MFOP) is different from traditional outcome prediction problem, its definition is given as follows:

**Definition 1 (Multifactorial outcome prediction (MFOP)):** *MFOP is defined as predicting outcomes through multiple modalities by using different classifiers in multi-criterion way so as to obtain accurate, reliable and interpretable results.*

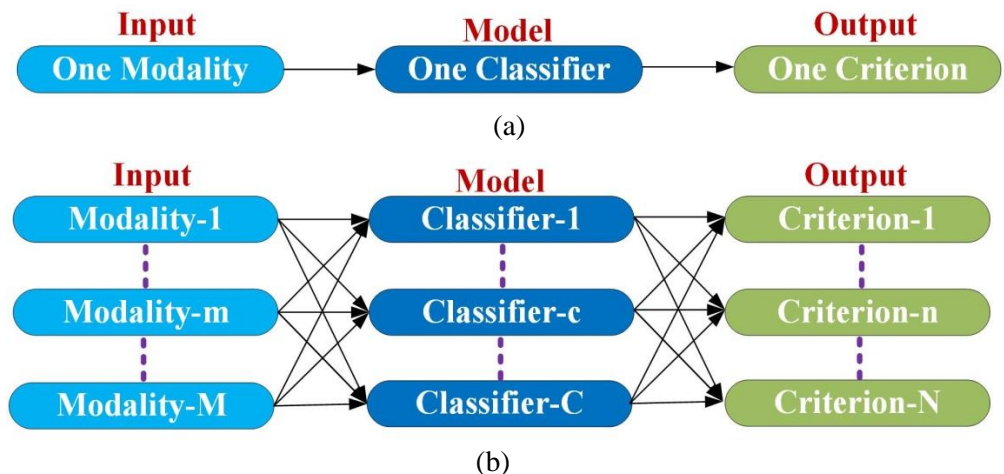

Fig. 1: The visualized difference between traditional outcome prediction and multifactorial outcome prediction (MFOP). In traditional outcome prediction, there is one modality, one classifier and one

criterion, while there are multiple modalities, multiple classifiers and multiple criterions in input, model construction and output in MFOP.

To facilitate understanding, the visualized differences between traditional outcome prediction and MFOP are shown in Fig. 1. For a typical traditional prediction problem (Fig. 1 (a)), there is one modality as input, one classifier for building the model, and one criterion as output, while MFOP (Fig. 2 (a) ) contains multiple modalities as input, multiple classifiers for constructing model, and multiple criterions. It can be seen that MFOP is a more generalized framework and the traditional outcome prediction is a special case of MFOP. Additionally, the three challenges (multi-modality, multi-classifiers and multi-criterion) are distributed in input, model construction and output in MFOP, which covers the most important three parts when building the predictive model.

The aim of MFOP not only obtains more accurate result which is to get the correct output label as much as possible, but also gets more reliable result which means maximizing the similarity between predicted output probability and true label vector. For example, assume that there are two models (A and B) for predicting the outcome and the true label vector is [1, 0]. The predicted output probability for model A is [0.9, 0.1], while the output for model B is [0.51, 0.49]. Based on the maximal probability output, the predicted label is same for the two models. However, since 0.9 is closer to 1 than 0.51, A is more reliable than B. Getting more reliable predictive result is of great importance when making the treatment plan. For example, when the output probability of distant failure is more than 0.75, the physician may determine adopt additional systemic therapy in lung SBRT. On the other hand, getting an interpretable results are also essential. Knowing which features are more important for prediction and how they are combined can be very powerful in helping physician understand and trust the whole decision procedure, leading to patient survival improvement.

## III. Method

### A. Overview

The overall framework of the M-radiomics is shown in Fig. 2, and its definition is as follows:

**Definition 2 (Multifaceted radiomics (M-radiomics)):** *M-radiomics is defined as extracting a large number of quantitative features from multi-modality imaging and analyzing through multiple classifiers based on multi-criteria.*

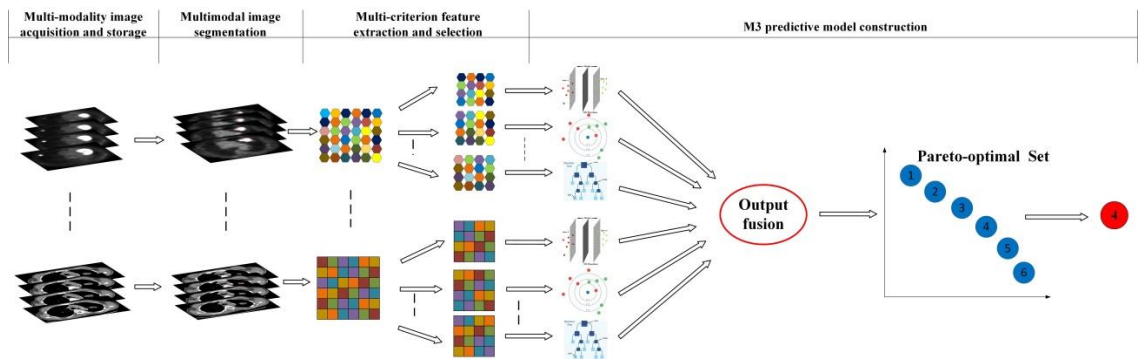

Fig. 2: The overall framework of M-radiomics. It consists of multimodal image segmentation, multi-criterion feature extraction and selection, M3 (multi-modality, multi-classifier and multi-objective) predictive model construction.

M-radiomics consists of multimodal image segmentation, multi-criterion feature extraction and selection, M3 (multi-modality, multi-classifier and multi-objective) predictive model construction. In the first step, the regions of interest (e.g. tumor) are segmented in a collaborative way in $M$ modalities. Then the quantitative features denoted by $fea^i = \{fea_1^i, fea_2^i, \cdots, fea_{M_i}^i\}, i = 1, \cdots, M$ are extracted from segmented region of interest, where $M_i$ represents the feature number for each modality. To achieve the accurate, reliable as well as balanced (sensitivity and specificity) performance and select the non-redundant features for each classifier, the multi-objective based feature selection is necessary. Assume that there are $O$ objective functions and the goal is to simultaneously maximize all the objective functions:

$$f_{fea}^i = \max_{fea^i}(f_1^i, \cdots, f_j^i, \cdots, f_O^i), i = 1, \cdots, M, j = 1, \cdots, O, \quad (1)$$

The selected features are denoted by $fea\_s^i = \{fea\_s_1^i, fea\_s_2^i, \cdots, fea\_s_{M_i}^i\}, i = 1, \cdots, M$. In the third step, the M3 predictive model will be constructed. Assume that there are $N$ classes output and $C$ modality-specific classifiers. When the selected features are fed into the modality-specific classifiers, we can obtain the output probability which is denoted by $P_c = \{p_c^1, \cdots, p_c^n, \cdots, p_c^N\}, c = 1, \cdots, C, \sum_{n=1}^{N} p_c^n = 1$. Assume that the relative weight and model parameters of each modality-specific classifier is denoted by $w_c, c = 1, \cdots, C, 0 \leq w_c \leq 1$ and $mp = \{mp_1, \cdots, mp_c, \cdots, mp_C\}$, respectively. To obtain more reliable results, reliability denote by $r_c, c = 1, \cdots, C, 0 \leq r_c \leq 1$ is introduced in validation set based reliable fusion (VRF). Hence, the final output probability $P_{fin} = \{p^1, \cdots p^c, \cdots, p^C\}$ is inferenced through VRF:

$$P_{fin} = VRF(P_c, w_c, r_c), c = 1, \cdots, C, \quad (2)$$

Furthermore, to obtain more accurate and reliable model, the parameter $mp$ and weight $w_c$ need to be trained. The method is same as feature selection process. Assume that there are $PO$ objective functions and the goal is to maximize the following function:

$$f^M = \max_{mp, w_c}(f_1^M, \cdots f_{po}^M, \cdots, f_{PO}^M), c = 1, \cdots, C. \quad (3)$$

where $f_{po}^M$ represents the objective function.

The details of VRF will be described in the following subsection. Additionally, to obtain more reliable predictive results, a reliable optimization model as well as a new recursive two stage hybrid optimization algorithm (RTSH) is developed in this work, which will be described in the following subsection. On the other hand, the whole M-radiomics model is also visible and interpretable. Although several individual classifiers (e.g. support vector machine (SVM)) are black-box modelling, they can be interpreted through extracting rules [23, 24].

**B. Validation set based reliable fusion strategy**

Reliable fusion is described as fusing the output probability of individual classifiers with both weight and reliability, which reliability is the ability to assess a given problem and weight is the relative importance to other information sources [6]. In this work, we calculate the reliability and weight based on the validation set.

Assume that there are $N$ individual classifiers denoted by $C = \{C_1, \cdots, C_N\}$ with $M$ classes, where $P_i = \{p_i^1, p_i^2, \cdots, p_i^M\}, i = 1, \cdots N$ is the output probability for a test sample $x$. The reliability and weight for each individual model are denoted by $r_i$ and $w_i, i = 1, \cdots, N$, and the validation set is denoted by $V$. The final output probability $P_f$ is obtained through analytic evidential reasoning (AER) rule [6], that is:

$$P_f = AER(P_i, w_i, r_i), i = 1, \cdots, N, \tag{4}$$

The brief description of evidential reasoning and the inference process of AER are shown in Appendix.

To achieve this goal, reliability and weight need to be determined. In VRF, reliability is determined by the similarity between the output probability of the test sample $x$ and output probabilities of $K$ nearest neighbors in validation set $V$, that is:

$$r_i = \begin{cases} 0 & \text{when } l_i \neq l_j; j = 1, \cdots, K \\ 1 & \text{when } l_i = l_j \wedge p_{l_j} = 1; j = 1, \cdots, K \\ 0 < r_i < 1 & \text{in other situations} \end{cases}, \tag{5}$$

where $l_i$ is the output labels of test sample $x$. $l_j$ are the output labels of $K$ nearest neighbor samples in $V$, and $p_{l_j}$ are the corresponding output probabilities.

Since $\sum_{m=1}^{M} p_i^m = 1$, the output probability for classifier $C_i$ is the probability density distribution. Therefore, calculating the reliability of the classifier $C_i$ for test sample $x$ is transformed into measuring the probability distribution similarity. Since dice coefficient is a commonly used probability distribution measuring method [25], it is used here. As it is hard to calculate the similarity measure directly, dissimilarity measure is calculated at first. Assume that output probability of test sample $x$ is denoted by $p_i^m$ and output probability of one validation sample $v_k, k = 1, \cdots, K$ in $K$ nearest neighbors is denoted by $p_{i_k}^m$. Based on the dice coefficient, the dissimilarity measure which is denoted by $D_{i,k}(x, v_k)$ between test sample $x$ and validation sample $v_k$ is calculated as:

$$D_{i,k}(x, v_k) = \frac{\sum_{m=1}^{M}\left(p_i^m - p_{i_k}^m\right)^2}{\sum_{m=1}^{M}(p_i^m)^2 + \sum_{m=1}^{M}\left(p_{i_k}^m\right)^2}, i = 1, \cdots, N; k = 1, \cdots, K, \tag{6}$$

where $i$ represents $ith$ classifier and $m$ represents the $mth$ class. $k$ is the $kth$ validation sample in in $K$ nearest validation samples. Based on the definition of reliability in equation (5), when $l_i = l_j \wedge p_{l_j} = 1; j = 1, \cdots, N, j \neq i, r_i = 1$. To satisfy this condition, $D_{i,k}(x, v_k)$ is modified as:

$$D_{i,k}(x, v_k) = \frac{\sum_{m=1}^{M}\left(p_i^m - p_{i_k}^m\right)^2 * \sum_{m=1}^{M}(1-p_i^l)^2 + \sum_{m=1}^{M}(1-p_i^l)^2}{\sum_{m=1}^{M}(p_i^m)^2 + \sum_{m=1}^{M}\left(p_{i_k}^m\right)^2 + \sum_{m=1}^{M}(1-p_i^l)^2 + \sum_{m=1}^{M}(1-p_i^l)^2}, \tag{7}$$

where $p_i^l$ is the maximal output probability for test sample $x$. Then $D_{i,k}(x, v_k)$ is equally transformed as:

$$D_{i,k}(x, v_k) = \frac{\left(\sum_{m=1}^{M}(p_i^m - p_{i_k}^m)^2 + 1\right) * \sum_{j=1}^{M}(1-p_i^l)^2}{\sum_{m=1}^{M}(p_i^m)^2 + \sum_{m=1}^{M}\left(p_{i_k}^m\right)^2 + 2\sum_{j=1}^{M}(1-p_i^l)^2}, \tag{8}$$

which deduces the similarity measure between the output probability of test sample $x$ and a validation sample denoted by:

$$S_{i,k}(x, v_k) = 1 - \frac{\left(\sum_{m=1}^{M}(p_i^m - p_{i_k}^m)^2 + 1\right) * \sum_{j=1}^{M}(1-p_i^l)^2}{\sum_{m=1}^{M}(p_i^m)^2 + \sum_{m=1}^{M}(p_{i_k}^m)^2 + 2\sum_{j=1}^{M}(1-p_i^l)^2}, \quad (9)$$

where $S_{i,k}(x, v_k)$ represents the similarity measure. Since there are $K$ validation samples, the overall similarity denoted by $S_i(x)$ is:

$$S_i(x) = \frac{1}{K}\sum_{k=1}^{K}\left(1 - \frac{\left(\sum_{m=1}^{M}(p_i^m - p_{i_k}^m)^2 + 1\right) * \sum_{j=1}^{M}(1-p_i^l)^2}{\sum_{m=1}^{M}(p_i^m)^2 + \sum_{m=1}^{M}(p_{i_k}^m)^2 + 2\sum_{j=1}^{M}(1-p_i^l)^2}\right), \quad (10)$$

Assume that $NV$ ($NV \leq K$) is the number of validation samples which the output labels are same as the test sample. To make sure the reliability equals 0 when the output labels of all the validation samples are different from the output label of test sample, We add $\frac{NV}{K}$ into $r_i(x)$, it is:

$$r_i(x) = \frac{NV}{K^2}\sum_{k=1}^{K}\left(1 - \frac{\left(\sum_{m=1}^{M}(p_i^m - p_{i_k}^m)^2 + 1\right) * \sum_{j=1}^{M}(1-p_i^l)^2}{\sum_{m=1}^{M}(p_i^m)^2 + \sum_{m=1}^{M}(p_{i_k}^m)^2 + 2\sum_{j=1}^{M}(1-p_i^l)^2}\right), i = 1, \cdots, N, \quad (11)$$

Similar to the reliability, the optimal weight $w_i$ is also determined by measuring the similarity between the output probability of test sample and the validation set. Since the only constraint for $w_i(x)$ is $0 \leq w_i \leq 1$, the dice coefficient can be directly used, that is:

$$w_i(x) = \sum_{k=1}^{K}\frac{2\sum_{m=1}^{M}p_i^m p_{i_k}^m}{\sum_{m=1}^{M}(p_i^m)^2 + \sum_{m=1}^{M}(p_i^m)^2}, , i = 1, \cdots, N, \quad (12)$$

Same with reliability, when the output labels of all the validation samples are different from the output label of test sample, we will let $w_i(x) = 0$. Therefore, we also add $\frac{NV}{K}$ into $w_i(x)$, that is:

$$w_i(x) = \frac{NV}{K}\sum_{k=1}^{K}\frac{2\sum_{m=1}^{M}p_i^m p_{i_k}^m}{\sum_{m=1}^{M}(p_i^m)^2 + \sum_{m=1}^{M}(p_i^m)^2}, , i = 1, \cdots, N, \quad (13)$$

The final output probability denoted by $p^m$ for test sample $x$ is calculated through AER by fusing the output probabilities of all the individual classifiers, that is:

$$p^m(x) = \frac{k\left[\prod_{i=1}^{N}\left(\frac{w_i p_i^m}{1+w_i-r_i} + \frac{1-r_i}{1+w_i-r_i}\right) - \prod_{i=1}^{N}\left(\frac{1-r_i}{1+w_i-r_i}\right)\right]}{1 - k\prod_{i=1}^{N}\left(\frac{1-r_i}{1+w_i-r_i}\right)}, m = 1, \cdots, M, \quad (14)$$

where $k$ is:

$$k = \left[\sum_{m=1}^{M}\left(\prod_{i=1}^{N}\left(\frac{w_i p_i^m}{1+w_i-r_i} + \frac{1-r_i}{1+w_i-r_i}\right)\right) - (M-1)\prod_{i=1}^{N}\left(\frac{1-r_i}{1+w_i-r_i}\right)\right]^{-1}. \quad (15)$$

Then the final label $L$ for test sample $x$ is determined by:

$$L(x) = \max(p^m(x)), m = 1, \cdots, M. \tag{16}$$

To show whether Equation (11) meets the definition of reliability, a numerical study is given as follows. Assume that there are four individual classifiers for a classification problem with three classes. The output probability of classifier $C_1$ is {0.8, 0.1, 0.1}. Table 1 shows the four group output probabilities for other three classifiers, where different groups represent different folders in a four-folder cross validation. Without loss of generality, it is assumed that the output probabilities for three models are same in all four groups. The reliabilities of $C_1$ in four groups are 0.7530, 0.8420, 0.9231 and 1.000 from Group1 to Group4. In general, the reliability gradually increases from Group1 to Group3 the outputs become more similar. On the other hand, although output probability is same in Group3, the reliability is still less than 1. Only when all the output probabilities from other classifiers are the same as (1.0, 0.0, 0.0), the reliability equals to 1.

Table 1: Output probabilities for three models in four groups.

| Model | Group1 | | | Group2 | | | Group3 | | | Group4 | | |
|---|---|---|---|---|---|---|---|---|---|---|---|---|
| C2 | 0.6 | 0.2 | 0.2 | 0.7 | 0.15 | 0.15 | 0.8 | 0.1 | 0.1 | 1.0 | 0.0 | 0.0 |
| C3 | 0.6 | 0.2 | 0.2 | 0.7 | 0.15 | 0.15 | 0.8 | 0.1 | 0.1 | 1.0 | 0.0 | 0.0 |
| C4 | 0.6 | 0.2 | 0.2 | 0.7 | 0.15 | 0.15 | 0.8 | 0.1 | 0.1 | 1.0 | 0.0 | 0.0 |

## C. Reliable Optimization model

Since the aim of multifactorial outcome prediction is not only to obtain more accurate results, but also get more reliable results, a reliable optimization model for training M-radiomics is desirable. Assume that the parameters in M-radiomics is denoted by $MP = \{MP_1, MP_2, \cdots, MP_P\}$. To obtain the balanced accurate results, both sensitivity and specificity are considered as the objective functions simultaneously. Assume that $f_{sen}$ and $f_{spe}$ represent the sensitivity and specificity objective function, respectively. They are defined as [3]:

$$f_{sen} = \frac{TP}{TP+FN}, \quad f_{spe} = \frac{TN}{TN+FP}, \tag{17}$$

where $TP$ is the number of true positives, $TN$ is the number of true negatives, $FP$ is the number of false positives, and $FN$ is the number of false negatives. The aim is to maximize the two objective functions simultaneously, that is:

$$f_{acc} = \max_{MP}(f_{sen}, f_{spe}), \tag{18}$$

where $f_{acc}$ represents the accuracy based objective function.

On the other hand, reliability based objective function denoted by $f_{rel}$ is also needed. Inspired by the sensitivity and specificity, we define the similarity based sensitivity and specificity, which are denoted by $f_{sim\_sen}, f_{sim\_spe}$, respectively. Assume that $P_{tp} = \{p_{tp}^1, p_{tp}^2, \cdots, p_{tp}^{TP}\}$ represents the probability output of true positives and the corresponding labelled vector is $T_{tp} = \{p_{tp}^1, p_{tp}^2, \cdots, p_{tp}^{TP}\}$. Based on the dice coefficient, the similarity measure of true positives $TP_{sim}$ is:

$$TP_{sim} = \sum_{k=1}^{TP} sim(p_{tp}^k, T_{tp}^k) = \sum_{k=1}^{TP} \frac{2\sum_{j=1}^{N} p_{tp,j}^k T_{tp,j}^k}{\sum_{j=1}^{N}(p_{tp,j}^k)^2 + \sum_{j=1}^{N}(T_{tp,j}^k)^2}, \quad (19)$$

Similarly, we can get the similarity measure of true negatives, false positives and false negatives, which are denoted by $TN_{sim}$, $FP_{sim}$ and $FN_{sim}$, respectively. Similar with the sensitivity and specificity, we can also obtain $f_{sim\_sen}, f_{sim\_spe}$, they are:

$$f_{sim\_sen} = \frac{TP_{sim}}{TP_{sim} + FN_{sim}}, \quad f_{sim\_spe} = \frac{TN_{sim}}{TN_{sim} + FP_{sim}}, \quad (20)$$

These two type objective functions constitute the reliability based objective function $f_{rel}$, that is:

$$f_{rel} = \max_{MP}\left(f_{sim_{sen}}, f_{sim_{spe}}\right). \quad (21)$$

Since $f_{acc}$ describes the label output and $f_{rel}$ shows the probability output, they are correlated. However, when performing multi-objective optimization, the objective functions have be conflict among each other. Therefore, instead of optimizing the two type objective functions, we train them alternatively and a new recursive two stage hybrid optimization algorithm (RTSH) is proposed. RTSH consists of two stages, which the first stage is to train the accuracy based objective function and the second stage is to maximize the reliability based objective function. For the first stage, our previous proposed iterative multi-objective immune algorithm (IMIA) [3] is adopted. To improve the reliability in the second stage, we design a similarity-based multi-objective optimization algorithm (SMO) [26]. These two stages are performed recursively until the algorithm achieves the maximal generation.

**D. Recursive two stage hybrid optimization algorithm**

Before describing the proposed RTCH, IMIA is briefly described as follows.

*Step 1* – Initialization: Model parameters are optimized directly as the value is continuous. We use $G_{max}$ to denote the maximum number of generations and $D(j) = \{d_1, \cdots, d_P\}$ denote the population, with P as the number of individuals in one population.

*Step 2* – Clonal operation: Proportional cloning will be used in this study[27]. To diversify the population, an individual with a larger crowding-distance is reproduced more times, and the clonal time $q_i$ for each individual is calculated as: $q_i = \left\lceil n_c \times \frac{\delta(d_i, D)}{\sum_{j=1}^{|A|} \delta(d_j, D)} \right\rceil$, where $n_c$ is the expectant value of the clonal population and $\lceil \ \rceil$ is the ceiling operator. $\delta(d_i, D)$ represents the crowding distance.

*Step 3* – Mutation operation: The mutation operation will be performed on the cloned population $C(j)$. For each locus in the individual, a random mutation probability $(MP_i)$ will be generated. If $MP_i$ is larger than a general mutation probability $(GMP_i)$, the mutation will occur. The mutated population is denoted by $M(j)$. $D(j)$ and $M(j)$ are combined to form a new population $F(j)$.

*Step 4* – Deleting operation: If there are duplicated solutions in the new population $F(j)$, we will only keep the unique one and delete other duplicated solutions. If the size of F(j) is less than P, step 2 should be used to generate more mutated individuals; otherwise, step 5 should be used. In contrast to other immune-inspired algorithms, this deleting operation is a new operator of the proposed IMIA. When the same solutions are generated in the Pareto-optimal set, the diversity of the individuals in a population will be reduced. We will perform the deleting operation to ensure that all the solutions in the Pareto-optimal set are different after executing the clonal and mutation operations.

*Step 5* – Updating population: The selected features and model parameters for each individual are taken as the input for model. Then, $f_{sen}$ and $f_{spe}$ are calculated through a cross-validation. The individual in all feasible solutions is sorted in the descending order using a fast non-dominated sorting approach[28] according to the AUC of each solution. We will obtain the Pareto-optimal set according to AUC because it is one of the most important criteria for the performance evaluation of a predictive model.

A Pareto-optimal solution set is generated after finish training. SMO is similar with IMIA in most steps except from step 2 and step 5. In step 2, a new similarity-based proportional cloning operation was proposed, where the solution with higher similarity was reproduced multiple times. Specifically, the clonal time $CLT_i$ for each solution is calculated as:

$$CLT_i = \left\lceil n_c \times \frac{sim(d_i)}{\sum_{i=1}^{H} sim(d_i)} \right\rceil, \tag{22}$$

where $n_c$ is the expected value of the clonal solution set and ⌈ ⌉ is the ceiling operator. The similarity measure for solution $d_i$ denoted by $sim(d_i)$ is calculated as:

$$sim(d_i) = \sum_{k=1}^{K} \frac{2 \sum_{j=1}^{N} P_j^k T_j^k}{\sum_{j=1}^{N} \left(P_j^k\right)^2 + \sum_{j=1}^{N} \left(T_j^k\right)^2}, \tag{23}$$

where $K$ is the number of training samples and $T_j^k$ is the label vector. In step 5, the solution in is sorted in the descending order using the fast non-dominated sorting approach according to reliability based objective function of each solution. The procedures of RTCH is shown in table 2, where $R$ is the maximal generation. After generating the Pareto-optimal solution set, the optimal solution is selected based on the area under the curve (AUC) and accuracy.

Table 2: Procedures of RTCH.

| |
|---|
| **Input**: Initial population, set $r = 0, j = 0$. |
| *while* $r < R$ |
|     Stage 1: *while* $j < G$, perform IMIA: |
|         Step 1: Proportional clonal operation. |
|         Step 2: Static mutation operation. |
|         Step 3: Deleting operation. |
|         Step 4: AUC based fast nondominated sorting approach for updating solution set. |
|     End *while* |
|     Stage 2: *while* $j < G$, perform SMO: |
|         Step 1: similarity-based proportional cloning operation. |
|         Step 2: Static mutation operation. |
|         Step 3: Deleting operation. |
|         Step 4: Similarity based fast nondominated sorting approach for updating solution set. |
|     End *while* |
| End *while* |

| Output: Pareto-optimal solution set. |
|---|

**D. Training and testing procedures**

In this subsection, the training and testing procedures of M-radiomics for handling two MFOP problems in H&N cancer is described. Since the tumor is segmented manually, the training part mainly consists of feature extraction, feature selection and predictive model construction. Totally 257 features including intensity, texture and geometry features are extracted for FDG-PET and CT images, respectively. When applying RTSH for multi-criterion selecting features, the binary initialization is adopted, which "1" represents the selected feature and "0" represents the unselected feature. During the M3 model construction, six different individual classifiers are used, including support vector machine, logistic regression (LR), discriminant analysis (DA), decision tree (DT), K-nearest-neighbor (KNN), and naive Bayesian (NB). Hence, there are 12 modality-specific individual classifiers.

The testing part also divides into three stages. For a test sample, first, the features for each modality-specific classifier are selected at first stage. In the second stage, each modality-specific classifier outputs a probability. At final stage, the final probability is obtained by combining all the output probabilities as well as reliability and weight using the VRF. Then, and the label with maximal output probability is determined as the final label output.

## IV. Experiments

**A. Dataset description**

For distant metastasis prediction in H&N cancer, FDG-PET and radiotherapy planning CT from 188 patients are used. All patients had pre-treatment FDG-PET/CT scans between April 2006 and November 2017. The follow up time ranges from 6 months to 112 months and the median follow up time is about 43 months. Sixteen percent (16%) of these patients had distant metastasis. All the image features are extracted from 3D tumor provided by TCIA. For predicting locoregional recurrence, FDG-PET and CT from 100 patients with definitive radiation therapy were retrospectively used. Among these patient data, 40 patients experienced locoregional recurrence.

**B. Setup**

When building the M-radiomics model for two clinical problems, four group experiments were designed in this study so that the effectiveness of M-radiomics as well as new RCF and RTSH can be comprehensively validated. They are: (1) multi-modality evaluation; (2). Multi-classifier evaluation; (3) validating RTSH; (4) new RCF (RCF-II) validation. All the experiments performed two-fold cross-validation. Sensitivity (SEN), specificity (SPE), area under the curve (AUC) and accuracy (ACC) were used to evaluate the performance. Additionally, the reliability is evaluated through the similarity based sensitivity and specificity.

In addition to two single modality (PET, CT) in the first experiment, the method which directly combined three modality features (named as DCM) was also compared with M-radiomics. To reduce the effect of other factors, only support vector machine (SVM) is used for modelling. Three modalities are used in our proposed way in the following three experiments. In the second experiment, six individual classifiers, including SVM, logistic regression (LR), discriminant analysis (DA), decision tree (DT), K-nearest-neighbor (KNN), and naive Bayesian (NB) were used for comparison. IMIA was used for optimizing in the above two experiments. In the third experiment, the single objective (SO) based optimization method [4] and IMIA were performed for comparison. The recursive time is set as 5 and the generation time in each recursion is set as 20 in RTSH. The population number as well as maximal generation was set as 100 in IMIA and SO, while the clonal factor was set as 200 and the mutation rate was set as 0.9 in IMIA. The previous RCF (RCF-I), evidential reasoning fusion (ERF) and weighted fusion (WF) were compared with RCF-II and RTSH is applied in the final experiment.

**C. Results and analysis for study 1**

The evaluation results on four modalities are shown in table 3. MR performs best in ACC and SEN, while DCM and Clinic obtains best ACC and SPE, respectively. It shows the combined models outperform single modality based models. MR also outperforms DCM in three criterions and only AUC is slightly lower, which means that the proposed combined method is more reasonable. Table 4 presents the results of six individual classifiers and combined model, among which MR performs best. This is because different classifiers improve the diversity of output, which makes the combined model be more robust. Three optimization algorithms are compared in the third experiment as shown in table 5 while the results of different fusion strategies shown in table 6. It shows that RTSF can obtain the best performance except from AUC as AUC is considered as the objective function in SO. RCF-II outperforms other three fusion strategies in all four criterions because of the advantages of new reliability calculation method and the way of optimizing relative weight.

**Table 3.** Evaluation results on five modalities.

| Modality | AUC | ACC | SEN | SPE |
|---|---|---|---|---|
| Clinic | 0.6804 | 0.7294 | 0.5333 | **0.7785** |
| PET | 0.6363 | 0.7243 | 0.5000 | 0.7658 |
| CT | 0.7094 | 0.6649 | 0.6333 | 0.6709 |
| DCM | **0.7181** | 0.6702 | 0.6333 | 0.6772 |
| MR | 0.7175 | **0.7347** | **0.6667** | 0.7595 |

**Table 4.** Evaluation results on six classifiers and combined results.

| Classifier | AUC | ACC | SEN | SPE |
|---|---|---|---|---|
| DA | 0.7264 | 0.6542 | 0.6333 | 0.6519 |
| DT | 0.6592 | 0.7181 | 0.5333 | **0.7532** |
| KNN | 0.7158 | 0.7287 | 0.6000 | 0.7341 |
| LR | 0.7124 | 0.6915 | **0.6667** | 0.6962 |
| NB | 0.6224 | 0.5957 | **0.6667** | 0.5886 |
| SVM | 0.7175 | 0.7347 | **0.6667** | 0.7395 |
| M-radiomics | **0.7340** | **0.7387** | **0.6667** | 0.7455 |

**Table 5.** Evaluation results on three optimization algorithms.

|  | AUC | ACC | SEN | SPE |
|---|---|---|---|---|
| SO | **0.7595** | 0.7074 | 0.6333 | 0.7215 |
| IMIA | 0.7340 | 0.7387 | **0.6667** | 0.7455 |
| RTSF | 0.7500 | **0.7478** | **0.6667** | **0.7595** |

**Table 6.** Evaluation results on four fusion strategies.

|  | AUC | ACC | SEN | SPE |
|---|---|---|---|---|
| WF | 0.7325 | 0.7074 | 0.6667 | 0.7125 |
| DSF | 0.7068 | 0.6649 | 0.6667 | 0.6646 |
| RCF-I | 0.7331 | 0.7340 | 0.6667 | 0.7435 |
| RCF-II | **0.7500** | **0.7478** | **0.6667** | **0.7595** |

**D. Results and analysis for study 2**

AUC, accuracy (ACC) sensitivity (SEN) and specificity (SPE) were taken as the evaluation criteria. Three modalities including PET, CT and PET & CT were used and two-folder cross-validation was performed. Table 7 shows the predictive results of three modalities, and table 8 shows the results of six individual classifier based radiomics and M-radiomics with PET&CT as input. Other than SEN, PET&CT modelling with M-radiomics obtains best performance in another three criteria.

Table 7: Predictive performance for three modalities.

| Modality | AUC | ACC | SEN | SPE |
|---|---|---|---|---|
| PET | 0.7473 | 0.7300 | **0.6500** | 0.7833 |
| CT | 0.7633 | 0.7500 | **0.6500** | 0.8167 |
| PET&CT | **0.7848** | **0.7800** | **0.6500** | **0.8667** |

Table 8: Predictive results for six individual classifiers and M-radiomics.

| classifier | AUC | ACC | SEN | SPE |
|---|---|---|---|---|
| SVM | 0.7308 | 0.7200 | **0.6500** | 0.7667 |
| LR | 0.7292 | 0.6700 | 0.6250 | 0.7000 |
| DA | 0.7129 | 0.7000 | 0.6000 | 0.7667 |
| DT | 0.7571 | 0.7300 | **0.6500** | 0.7833 |
| KNN | 0.7413 | 0.7100 | 0.5500 | 0.8167 |
| NB | 0.7173 | 0.7300 | 0.6000 | 0.8167 |
| M-radiomics | **0.7848** | **0.7800** | **0.6500** | **0.8667** |

## VI. Conclusions

In this work, a new problem termed as multifactorial outcome prediction (MFOP) was proposed. Three mainly challenges including multi-modality, multi-classifier and multi-criterion are integrated into a unified framework. The aim of MFOP is to obtain more accurate, more reliable and interpretable predictive results.

To handle the new MFOP, a new multifaceted radiomics (M-radiomics) model was proposed. M-radiomics consists of three parts, they are multimodal image segmentation, multi-criterion feature extraction and selection, M3 (multi-modality, multi-classifier and multi-objective) predictive model construction. The above three challenges are handled very well through M-radiomics.

We also developed a validation set based reliable fusion strategy (VRF) and a reliable optimization model in M-radiomics so as to improve the reliability of predictive results. In VRF, reliability as well as weight is introduced, and they were calculated based on the similarity measure between the output probability of test sample and validation sample set. In reliable optimization model, similarity based sensitivity and specificity were introduced to maximize the reliability, and a new recursive two stage hybrid optimization algorithm (RTSH) was proposed. This two stage optimization algorithm can ensure that M-radiomics can obtain accurate and reliable predictive results.

Two clinical problems including distant metastasis and locoregional recurrence prediction in H&N cancer were modelled through M-radiomics. The experimental results demonstrated that M-radiomics model outperformed current typical radiomics models. Compared with other fusion strategies and optimization algorithms, the proposed VRF and RTSH can obtain more reliable predictive results.

## Acknowledgements

This work is supported by the American Cancer Society (ACS-IRG-02-196) and the US National Institutes of Health (R01 EB020366).

## Appendix

### A. Brief description of evidential reasoning rule

Assume that $\Theta = \{h_1, h_2, \cdots, h_H\}$ is a set of mutually exclusive and collectively exhaustive hypotheses, where $\Theta$ is referred to as a frame of discernment. The power set of $\Theta$ consists of all its subsets, denoted by $P(\Theta)$ or $2^\Theta$, as follows:

$$P(\Theta) = 2^\Theta = \{\emptyset, \{h_1\}, \cdots, \{h_H\}, \{h_1, h_2\}, \cdots, \{h_1, h_M\}, \cdots, \{h_1, h_{H-1}\}, \Theta\}, \tag{A.1}$$

where $\{h_1, h_2\}, \cdots, \{h_1, h_M\}, \cdots, \{h_1, h_{H-1}\}$ are the local ignorance. In the ER rule, a piece of evidence $e_i$ is represented as a random set and profiled by a belief distribution (*BD*), as:

$$e_i = \{(\theta, p_{\theta,i}), \forall \theta \subseteq \Theta, \sum_{\theta \subseteq \Theta} p_{\theta,i} = 1\}, \tag{A.2}$$

where $(\theta, p_{\theta,i})$ is an element of evidence $e_i$, indicating that the evidence points to proposition $\theta$, which can be any subset of $\Theta$ or any element of $P(\Theta)$ except from the empty set, to the degree of $p_{\theta,i}$ referred to as probability or degree of belief, in general. $(\theta, p_{\theta,i})$ is referred to as a focal element of $e_i$ if $p_{\theta,i} > 0$.

The reasoning process in the ER rule is performed by defining a weighted belief distribution with reliability (*WBDR*) [29]:

$$m_i = \{(\theta, \widetilde{m}_{\theta,i}), \forall \theta \subseteq \Theta; (P(\Theta), \widetilde{m}_{P(\Theta),i}\}, \tag{A.3}$$

where $\widetilde{m}_{\theta,i}$ measures the degree of support for $\theta$ from $e_i$ with both weight and reliability being taken into account, defined as follows:

$$\widetilde{m}_{\theta,i} = \begin{cases} 0, & \theta = \emptyset \\ c_{rw,i} m_{\theta,i}, & \theta \subseteq \Theta, \theta \neq \emptyset \\ c_{rw,i}(1 - r_i), & \theta = P(\Theta) \end{cases} \quad or \quad \widetilde{m}_{\theta,i} = \begin{cases} 0, & \theta = \emptyset \\ \widetilde{w}_i p_{\theta,i}, & \theta \subseteq \Theta, \theta \neq \emptyset \\ 1 - \widetilde{w}_i, & \theta = P(\Theta) \end{cases}. \tag{A.4}$$

$m_{\theta,i} = w_i p_{\theta,i}$ and $c_{rw,i} = 1/(1 + w_i - r_i)$ is a normalization factor, which satisfies $\sum_{\theta \subseteq \Theta} \widetilde{m}_{\theta,i} + \widetilde{m}_{P(\Theta),i} = 1$. $\widetilde{w}_i = c_{rw,i} w_i$, is acting as a new weight. Then the ER rule combines multiple pieces of evidence recursively. If two pieces of evidence $e_1$ and $e_2$ are independent, $e_1$ and $e_2$ jointly support proposition $\theta$ denoted by $p_{\theta,e(2)}$, which is generated as follows:

$$p_{\theta,e(2)} = \begin{cases} 0 & \theta = \emptyset \\ \frac{\widehat{m}_{\theta,e(2)}}{\sum_{D \subseteq \Theta} \widehat{m}_{D,e(2)}} & \theta \subseteq \Theta, \theta \neq \emptyset \end{cases}' \tag{A.5}$$

$$\widehat{m}_{\theta,e(2)} = \left[(1 - r_2) m_{\theta,1} + (1 - r_1) m_{\theta,2}\right] + \sum_{B \cap C = \theta} m_{B,1} m_{C,2} \qquad \forall \theta \subseteq \Theta, \tag{A.6}$$

When there are $L$ pieces of independent evidence, the jointly support proposition $\theta$ denoted by $\widehat{m}_{\theta,e(L)}$ can be generated by the following two equations:

$$\widehat{m}_{\theta,e(L)} = \left[(1 - r_i) m_{\theta,e(i-1)} + m_{P(\Theta),e(i-1)} m_{\theta,i}\right] + \sum_{B \cap C = \theta} m_{B,e(i-1)} m_{C,i}, \quad \forall \theta \subseteq \Theta, \tag{A.7}$$

$$\widehat{m}_{P(\Theta),e(L)} = (1 - r_i) m_{P(\Theta),e(i-1)}, \tag{A.8}$$

After obtaining the normalization, the combined $BD$ $p_\theta$ can be calculated by the following equation:

$$p_\theta = \frac{\widehat{m}_{\theta,e(L)}}{1 - \widehat{m}_{P(\Theta),e(L)}}, \quad \forall \theta \subseteq \Theta. \tag{A.9}$$

**B. Inference of analytic evidential reasoning rule**

As there's no local ignorance in outcome prediction, they are pruned in ER rule. Under no local ignorance, the $BD$ for each evidence $e_i$ is reduced to the following format:

$$e_i = \{(\theta_h, p_{h,i}), h = 1, \cdots, H; \sum_{\theta_h = 1}^M p_{h,i} = 1\}, i = 1, \cdots, N \tag{B.1}$$

And *WBDR* is reduced to:

$$m_i = \{(\theta_h, \widetilde{w}_i p_{h,i}), h = 1, \cdots, H; (P_i(\Theta), (1 - \widetilde{w}_i)\}, i = 1, \cdots, N \tag{B.2}$$

where $\theta_h$ is the class and $p_{h,i}$ is the corresponding output score of individual classifier $i$. $\widetilde{w}_i$ is the new weight.

Since normalization in the evidence combination can be applied at the end of the process without changing the combination result, we do not consider normalization when combining all the evidence but apply it in the end. Assume that $\widehat{m}_{\theta_h,l}, h = 1, \cdots, H$ and $\widehat{m}_{P(\Theta),l}$ denote the *WBDR* generated by combining the first $l$ evidence. We first consider a condition of $l = 2$: the combination of two evidences

(output scores from two classifiers) without normalization. The combined *WBDR* generated by aggregating the two evidences by orthogonal sum operation are given as follows.

$$\widehat{m}_{\theta_h,2} = m_{\theta_h,1} m_{P(\Theta),2} + m_{\theta_h,2} m_{P(\Theta),1} + m_{\theta_h,1} m_{\theta_h,2} \tag{B.3}$$

$$= m_{\theta_h,1}(m_{\theta_h,2} + m_{P(\Theta),2}) + m_{\theta_h,2} m_{P(\Theta),1}$$

$$= m_{\theta_h,1}(m_{\theta_h,2} + m_{P(\Theta),2}) + m_{P(\Theta),1}(m_{\theta_h,2} + m_{P(\Theta),2}) - m_{P(\Theta),1} m_{P(\Theta),2}$$

$$= (m_{\theta_h,1} + m_{P(\Theta),1})(m_{\theta_h,2} + m_{P(\Theta),2}) - m_{P(\Theta),1} m_{P(\Theta),2}$$

$$= \prod_{i=1}^{2}(m_{\theta_h,i} + m_{P(\Theta),i}) - \prod_{i=1}^{2} m_{P(\Theta),i},$$

And

$$\widehat{m}_{P(\Theta),2} = \prod_{i=1}^{2} m_{P(\Theta),i}, \tag{B.4}$$

Assume that the following equations are true for the (*l*-1) evidences. Let $l_1 = l - 1$ and:

$$\widehat{m}_{\theta_h,l_1} = \prod_{i=1}^{l-1}(m_{\theta_h,i} + m_{P(\Theta),i}) - \prod_{i=1}^{l-1} m_{P(\Theta),i}, \tag{B.5}$$

$$\widehat{m}_{P(\Theta),l_1} = \prod_{i=1}^{l-1} m_{P(\Theta),i}, \tag{B.6}$$

The above combined probability masses are further aggregated with the *l*th evidence. The combined probability masses are given as:

$$\widehat{m}_{\theta,l} = \widehat{m}_{\theta_h,l_1} m_{\theta_h,l} + \widehat{m}_{\theta_h,l_1} m_{P(\Theta),l} + m_{\theta_h,l} \widehat{m}_{P(\Theta),l_1} \tag{B.7}$$

$$= \widehat{m}_{\theta_h,l_1}(m_{\theta_h,l} + m_{P(\Theta),l}) + m_{\theta_h,l} \widehat{m}_{P(\Theta),l_1}$$

$$= \widehat{m}_{\theta_h,l_1}(m_{\theta_h,l} + m_{P(\Theta),l}) + \widehat{m}_{P(\Theta),l_1}(m_{\theta_h,l} + m_{P(\Theta),l}) - \widehat{m}_{P(\Theta),l_1} m_{P(\Theta),l}$$

$$= (\widehat{m}_{\theta_h,l_1} + \widehat{m}_{P(\Theta),l_1})(m_{\theta_h,l} + m_{P(\Theta),l}) - \widehat{m}_{P(\Theta),l_1} m_{P(\Theta),l}$$

$$= (m_{\theta_h,l} + m_{P(\Theta),l})((\prod_{i=1}^{l-1}(m_{\theta_h,i} + m_{P(\Theta),i}) - \prod_{i=1}^{l-1} m_{P(\Theta),i}) + \prod_{i=1}^{l-1} m_{P(\Theta),i}) -$$
$$m_{P(\Theta),l} \prod_{i=1}^{l-1} m_{P(\Theta),i}$$

$$= \prod_{i=1}^{l}(m_{\theta_h,i} + m_{P(\Theta),i}) - \prod_{i=1}^{l} m_{P(\Theta),i},$$

And

$$\hat{m}_{P(\Theta),l} = m_{P(\Theta),l}\hat{m}_{P(\Theta),l_1} = m_{P(\Theta),l}\prod_{i=1}^{l-1} m_{P(\Theta),i} = \prod_{i=1}^{l} m_{P(\Theta),i}, \tag{B.8}$$

Then we normalize the combined *WBDR* results. Assume that $k$ is the normalization factor, therefore

$$k\left(\sum_{h=1}^{H}\hat{m}_{\theta_h,l} + \hat{m}_{P(\Theta),l}\right) = 1, \tag{B.9}$$

That is:

$$k\left(\sum_{h=1}^{H}\left(\prod_{i=1}^{l}(m_{\theta_h,i} + m_{P(\Theta),i}) - \prod_{i=1}^{l} m_{P(\Theta),i}\right) + \prod_{i=1}^{l} m_{P(\Theta),i}\right) = 1, \tag{B.10}$$

$$k\left(\sum_{h=1}^{H}\left(\prod_{i=1}^{l}(m_{\theta_h,i} + m_{P(\Theta),i})\right)\right) - k(H-1)\prod_{i=1}^{l} m_{P(\Theta),i} = 1, \tag{B.11}$$

So

$$k = \left(\sum_{h=1}^{H}\left(\prod_{i=1}^{l}(m_{\theta_h,i} + m_{P(\Theta),i})\right) - (H-1)\prod_{i=1}^{l} m_{P(\Theta),i}\right)^{-1}, \tag{B.12}$$

Therefore,

$$m_{\theta,l} = k\hat{m}_{\theta,l}, \quad m_{P(\Theta),l} = k\hat{m}_{P(\Theta),l}, \tag{B.13}$$

where $m_{\theta,l}$ and $m_{P(\Theta),l}$ are the combined *WBDR* after normalization. So the *BD* $p_m$ after combining $l$ evidence is:

$$p_h = \frac{m_{\theta,l}}{1 - m_{P(\Theta),l}} = \frac{k\hat{m}_{\theta,l}}{1 - k\hat{m}_{P(\Theta),l}} = \frac{k\left(\prod_{i=1}^{l}(m_{\theta,i} + m_{P(\Theta),i}) - \prod_{i=1}^{l} m_{P(\Theta),i}\right)}{1 - k\prod_{i=1}^{l} m_{P(\Theta),i}}, h = 1,\cdots,H, \tag{B.14}$$

Based on Eq. (A.4) and $l = N$, the final *BD* is:

$$p_h = \frac{k\left[\prod_{i=1}^{N}\left(\frac{w_i p_{h,i}}{1+w_i-r_i} + \frac{1-r_i}{1+w_i-r_i}\right) - \prod_{i=1}^{N}\left(\frac{1-r_i}{1+w_i-r_i}\right)\right]}{1 - k\prod_{i=1}^{N}\left(\frac{1-r_i}{1+w_i-r_i}\right)}, h = 1,\cdots,H, \tag{B.15}$$

and $k$ is:

$$k = \left[\sum_{h=1}^{H}\left(\prod_{i=1}^{N}\left(\frac{w_i p_{h,i}}{1+w_i-r_i} + \frac{1-r_i}{1+w_i-r_i}\right)\right) - (H-1)\prod_{i=1}^{N}\left(\frac{1-r_i}{1+w_i-r_i}\right)\right]^{-1}. \tag{B.16}$$